\documentclass[twocolumn,english,aps,pra,showpacs]{revtex4}
\usepackage{amssymb}
\usepackage{amsmath}
\usepackage{colordvi}
\usepackage{amsthm}
\usepackage{epsfig}
\usepackage{graphicx}
\usepackage{subfigure}
\usepackage{color}

\def\avg(#1){\langle#1\rangle}

\def\be{\begin{equation}}
\def\ee{\end{equation}}
\def\bea{\begin{eqnarray}}
\def\eea{\end{eqnarray}}

\begin{document}

\title{\textbf{Quantum phase transition of nonlocal Ising chain with transverse
field in a resonator}}
\author{Yu-Na Zhang$^{\text{1},{2}}$}
\author{Xi-Wang Luo$^{\text{1},{2}}$}
\author{Guang-Can Guo$^{\text{1},{2}}$}
\author{Zheng-Wei Zhou$^{\text{1},{2}}$}
\email{zwzhou@ustc.edu.cn}
\author{Xingxiang Zhou$^{\text{1},{2}}$}
\email{xizhou@ustc.edu.cn}
\address{$^{\text{1}}$Key Laboratory of Quantum Information,
University of Science and \\ Technology of China, Hefei, Anhui
230026, P. R. China\\ $^{\text{2}}$Synergetic Innovation Center
of Quantum Information and Quantum Physics, University of Science
and Technology of China, Hefei, Anhui 230026, China}

\begin{abstract}

We study the quantum phase transition in a spin chain with variable Ising
interaction and position-dependent coupling to a resonator field. Such a
complicated model, usually not present in natural physical systems, can be
simulated by an array of qubits based on man-made devices and
exhibits interesting behavior. We show that, when the coupling between
the qubit and field is strong enough, a super-radiant phase transition occurs,
and it is possible to pick a particular field mode to undergo this phase
transition by properly modulating the strength of the Ising interaction. We also
study the impact of the resonator field on the magnetic properties of the spin
chain, and find a rich set of phases characterized by distinctive qubit
correlation functions.

\end{abstract}


\maketitle

\section{Introduction}
Quantum simulation is a powerful tool to study difficult physics problems
that cannot be easily solved analytically or simulated with a classical
computer \cite{R.P., S.L., B.N., J.I., H.H., P.H.}. In order to study such hard problems, the
simulation system must be carefully designed and set up to capture as much
as possible essence of the simulated problem. This requirement often
poses a great challenge and can only be met to a certain degree. It is one
of the main reasons why many quantum simulation protocols are very difficult to
realize experimentally. This issue is especially prominent in simulation
systems based on artificial atoms such as Josephson devices, because many of
their properties are fundamentally different than those of natural physical
particles \cite{Y.G., J.F., M.R., L.T., L.T.2}.

Though the inevitable discrepancy between the simulation and simulated systems
is often considered an obstacle in quantum simulation, it can also provide
opportunities for studying physics models under conditions not easily
accessible in natual physical systems. This is because, due to the excellent
controllability available in simulation systems based on man-made devices, one
can often tune the
critical physical parameters in a range far greater than what is feasible in a
natural physical system, and even realize configurations not possible in a
natural physical system. To explore such opportunities in quantum simulation
that have not received sufficient attention, we study the quantum phase
transitions in a nonlocal Ising chain interacting with a resonator. This problem
has its root in the well-known Dicke model \cite{R.H.,K.E.,Y.F.}, in which a
collection of identical and non-interacting two-level atoms are coupled to a
single electromagnetic(EM) field mode. Our problem has a few important
differences from the original Dicke model, in that many qubits are spread out
within a single wavelength of a multi-mode EM field, and there are
controllable interactions between the qubits. It is also possible to tune the
transverse field of each qubit individually. These characteristics, usually not
present in a natural atom-cavity system, are accessible in
simulation systems based on artificial atoms and man-made devices, and they have
profound impact on the behavior of the system and the method we use to treat it.

\begin{figure}
\includegraphics[width=1.0\linewidth]{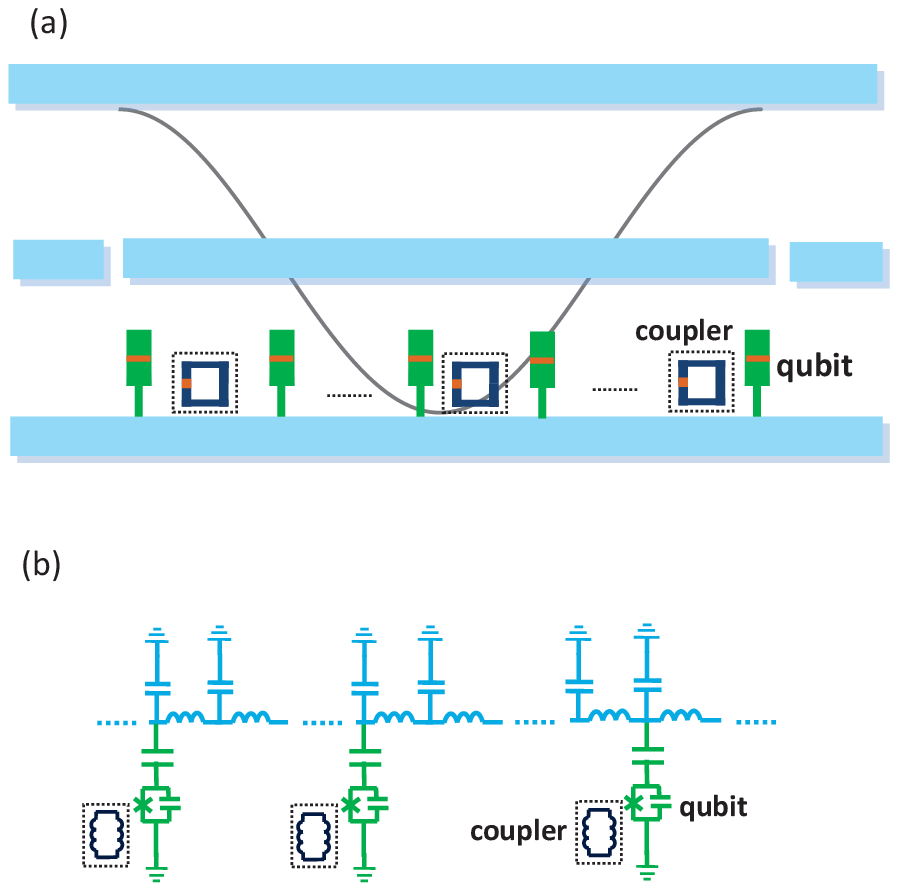}
\caption{(Color online) \footnotesize{(a) An array of charge qubits
capacitively coupled to the TLR.
The TLR consists of a center conductor and two ground planes. The voltage
between the center conductor and the ground planes is position-dependent,
as indicated by the cosine curve in the figure. The charge qubits located between the
center conductor and ground plane are capacitively coupled to the center
conductor.
The nearest-neighbor interaction between charge qubits is realized
by an rf-SQUID (in the dashed box).}
(b) The equivalent distributed circuit of (a).
}
\label{Fig:model}
\end{figure}

A possible physical realization of our simulation model is depicted in Fig.
\ref{Fig:model}. It consists of $N$ superconducting charge qubits placed at
equal distances and capacitively coupled to a transmission line resonator (TLR).
The charge qubit is biased at the charge degeneracy point to make it an
effective two-level system. The TLR supports mutiple resonant modes that the
charge qubits interact with \cite{A.B.,A.J.,P.C.}. This simulation system is analogous to
an atom-cavity system in the Dicke model, but with notable differences. In a
natural atom-cavity system, because the size of an atom is so small, the
displacement between individual atoms in an atomic cloud is negligible compared
with the wavelength of the EM field, and we can use the ``long-wavelength
approximation'' which assumes that the atoms are at the same
location and their coupling strengths to the EM field are identical. This
approximation does not apply in the system in Fig. \ref{Fig:model} because the
charge qubits, being macroscopic devices much larger than atoms, can spread
out along the entire TLR length which is also the wavelength (or its multiple)
of the EM modes that they couple to, and the position dependence of the coupling
strength must be taken into account. Another important distinction concerns the
fact that it is difficult to induce significant interactions between
charge-neutral atoms. This limitation can be overcome in our system by
introducing
coupling circuitry as shown in Fig. \ref{Fig:model}. By using large Josephson
junctions inductively coupled to the
charge qubits, we can induce strong and adjustable interactions between them,
greatly enriching the physics of our system.

In the following, we focus on phase transitions in our system of spatially
separated and interacting qubits coupled to a resonator field with multiple
modes. In the traditional Dicke model, the atom-field coupled system is subject to an instability due to the interaction between the atom and field. When the interaction
strength grows above a critical value, the field of the ground state of the system is
no longer in the vacuum mode. It becomes macroscopically occupied with photons and the system enters the so-called super-radiant phase \cite{C.E.}. In our system,
we find that the super-radiant phase transition can also occur when the
coupling strength between the qubits and the resonator field is strong enough, though
the details of the phase transition is much more complicated.
Further, by periodically modulating the strength of the interaction between
the qubits, we
can select which mode of the resonator field undergo the super-radiant phase
transition. We then study the magnetic properties of the qubit chain and its
phase transitions which are impacted by the state of the resonator field.

\section{model hamiltonian}

The full Hamiltonian of our system reads
\begin{equation}
 H=H_{Q}+H_R+H_{R-R}+H_{Q-R}+H_{Q-Q}.
\label{eq:Ht}
\end{equation}
Among these terms, the Hamiltonian of the $N$-qubit system
\begin{equation}
 H_Q=-\frac{E_z}{2}\sum_{j=0}^{N-1}\sigma_j^z
\end{equation}
is written in the eigenbases $\{(|0\rangle\pm|1\rangle)/\sqrt{2}\}$
at the charge degeneracy point, with $|0\rangle$ and $|1\rangle$ the 0 and 1 excess charge state. The multi-mode (labeled by the energy quantum number $l$) resonator field
Hamiltonian is
\begin{equation}
 H_{R}=\sum_{l}\omega_{l}b^{\dag}_lb_l.
\end{equation}
The coupling between the qubit system and the resonator field is assumed to be
dipolar and described by
\begin{equation}
 H_{Q-R}=-\sum_j\sum_l\frac{\lambda_l(j)}{\sqrt{N}}\sigma_j^x(b_l^\dag+b_l),
\end{equation}
where the coupling strength
\begin{equation}
 \lambda_l(j)=\lambda_0\sqrt{l}\cos(l{\pi}j/N)
\label{eq:lambda_pos_dep}
\end{equation}
is dependent on the position of the qubits which are assumed to be equally
spaced.
In addition, we also have terms for the self energy of the resonator field
and nearest-neighbor qubit interaction,
\begin{equation}
 H_{R-R}=\sum_lD_l(b_l+b_l^\dag)^2
\end{equation}
and
\begin{equation}
 H_{Q-Q}=-\sum_jJ(j)\sigma_j^y\sigma_{j+1}^y.
\end{equation}
Here, $D_l$ is the field self interaction strength
and $J(j)$'s characterize the position dependent Ising interaction
strength.

The Hamiltonian in Eq. (\ref{eq:Ht}) has several important differences
from the conventional Dicke model:
\begin{itemize}
 \item The EM field can have multiple modes, consistent with the situation in
physical resonators.
  \item The long wave approximation does not apply and the coupling between the
qubit and field is dependent on the position of the qubit.
  \item There is Ising interaction between nearest-neighbor qubits.
\end{itemize}
These new elements in our model have a profound impact on the system behavior
and phase transitions. As mentioned in the introduction, such a model
Hamiltonian can be realized using man-made devices with excellent
controllability such as the charge qubit - TLR system in Fig. \ref{Fig:model}.
As shown in Appendix A, in such a system the qubit energy is equal
to the Josephson energy $E_z$ of the charge qubit. The resonator mode
frequencies are determined by the parameters of the TLR,
$\omega_l=l\pi/d\sqrt{L_0C_0}$, $L_0$ and $C_0$ the inductance and capacitance
per unit length of the TLR and $d$ its length. The qubit-field coupling strength
$\lambda_0 = \frac{eC_g}{C_{\Sigma}}\sqrt{\frac{N\omega_1}{dC_0}}$, where $C_g$
and $C_\Sigma$ are the gate capacitance of the charge qubit and total
capacitance of the charge island, and the field self interaction strength
$D_l=\frac{1}{N}\sum_j\frac{C_\Sigma}{2e^2}\lambda_l^2(j)$.

\section{quantum phase transition}

\subsection{Mean field treatment}
\label{Subsect:mean_field}
The solution of our system is complicated by the fact that the resonator field
has multiple modes. To avoid nonessential complications and focus on the study
of phase transitions, we will adjust the system parameters such that no
more than one resonator mode has macroscopic occupation. To find the conditions
for such a setup, we consider one resonator mode $l$ first and use the
mean field approximation to simplify the qubit - resonator coupling term
$H_{Q-R}$ as
\begin{equation}
\begin{split}
\sum_j\frac{\lambda_{l}(j)}{\sqrt{N}}(b_{l}^{\dag}+b_{l})
\sigma_j^x=&\sum_j
2\phi_{l}\lambda_{l}(j)\sigma_j^x+\sqrt{N}(b_{l}^{\dag}
+b_{l})\Sigma^x_l\\
&-2N\phi_{l}\Sigma^x_l,
\end{split}
\label{Eq:mean_field_coupling}
\end{equation}
where the order parameters
\begin{equation}\label{eq:meanphi-S}
\begin{split}
\phi_{l}=&\langle{G}|\frac{(b_{l}^{\dag}+b_{l})}{2\sqrt{N}}|G\rangle\\
\Sigma^x_l=&\langle{G}|\frac{\sum_j\lambda_{l}(j)\sigma_j^x}{N}|G\rangle
\end{split}
\end{equation}
with the ground state of the system $|G\rangle$.

Under the mean field approximation, the qubit part of the Hamiltonian becomes
that of a nonlocal Ising chain with a transverse field dependent on
$\phi_{l}$, the order parameter for the resonator field. As shown in
Appendix B, it can be solved by the Jordan-Wigner transformation which maps
the Ising chain to a collection of fermionic quasiparticles. The energy per
particle for the system is
\begin{equation}\label{eq:singleparticle}
e_g=\bigg(\omega_{l}\phi_{l}^2+4D_{l}\phi^2_{l}
-\frac{1}{2N}\sum_{k}\Lambda_{k}(\phi_l)\bigg),
\end{equation}
where the spectrum of the quasiparticles $\Lambda_{k}$ is a complicated
function dependent on $\phi_l$ as shown in Appendix B. By
finding the value of $\phi_l$ that minimizes $e_g$, we can determine the
ground state energy and the order parameter $\phi_l^g$ for the resonator field.
Unlike in conventional Dicke problems, this problem cannot be solved
analytically because of the complicated quasiparticle spectrum $\Lambda_{k}$.
Thus we numerically solve for $\phi_{l}^g$,
the order parameter of the resonator field for the ground state.

\subsection{The super-radiant phase transition}

\begin{figure}
\includegraphics[width=1.0\linewidth]{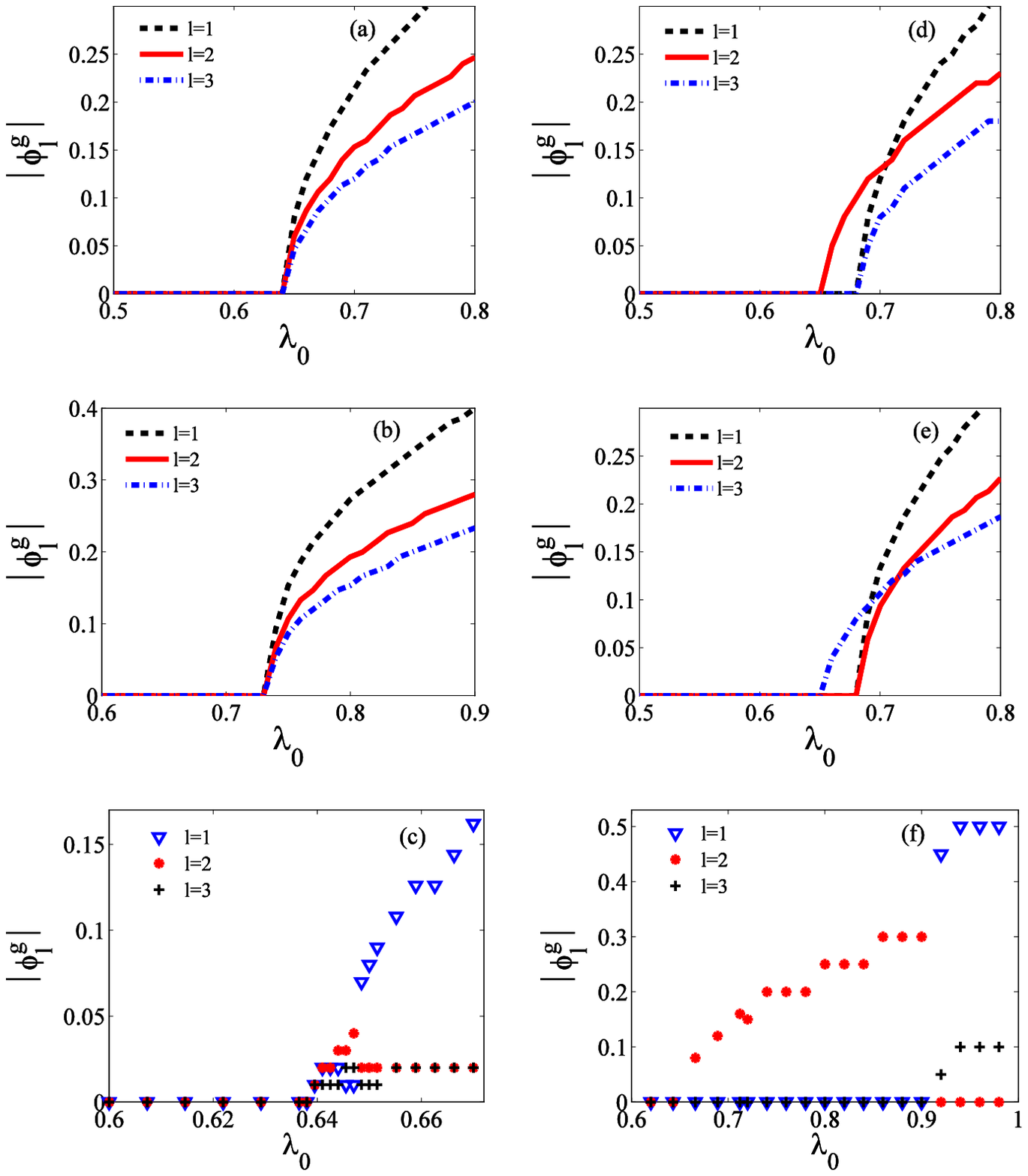}
\caption{(Color online) $\phi_l^g$ versus $\lambda_0$ for different resonator mode $l$.
In (a) and (b), only one mode is considered in the calculation, though a
different mode is used for each curve. the Ising interaction is
homogeneous, $J(j)=J$, and $J=0.05$ and $0.35$
respectively. In (c), the ground state values of three
modes ($l=1,2,3$) are plotted for homogeneous Ising interaction $J=0.05$
by considering all modes simultaneously in the calculation. In (d) and (e),
$J(j)$ has a rectangular waveform as in Eq. (\ref{eq:J1}) and Eq.
(\ref{eq:J2}), and $J_{max}=0.35$, $J_{min}=0.05$ are used. The mode $l=2$
and $l=3$ are
singled out since the critical value $\lambda_{0}^c$ for them to undergo the
super-radiant transition is the lowest. Only one mode is considered in each
calculation. In (f), three modes are considered simultaneously
in the calculation. $J(j)$ has a rectangular waveform as in Eq. (\ref{eq:J1}),
and $J_{max}=0.35$, $J_{min}=0.05$. In (a)-(f), the qubit biases,
$\frac{e^2}{2C_\Sigma}=8$ and $E_{z}=0.8$, are chosen to be accessible values in
typical experiments. The size of the system is $N=200$.}
\label{Fig:phimin}
\end{figure}

In this section, we focus on the state of the resonator field.
We start with the simple case of homogeneous Ising interaction $J(j)=J,\forall
j$, and calculate the ground state field order parameter $\phi_{l}^g$ for
different values of qubit-field coupling strength $\lambda_0$ and Ising
interaction strength $J$. In Fig. \ref{Fig:phimin}(a) and (b), the numerical
results of $\phi_{l}^g$ for $J=0.05$ and $J=0.35$ (in unit of $\omega_1$)
are shown.  It is seen that, when the qubit-field coupling $\lambda_0$ is
small, the ground state energy is minimized when $\phi^g_{l}=0$. When
$\lambda_0$ is greater than a critical value $\lambda_0^c$, $\phi^g_{l}$
becomes nonzero, indicating that the photon field has a
macroscopic occupation. Therefore, a super-radiant phase transition occurs
when the qubit-field coupling becomes strong enough.

We also calculated $\phi_{l}^g$ for different resonator modes $l$. We
find that, for homogeneous Ising interaction $J(j)=J$, the critical points
$\lambda_{0}^c$ for all resonator modes are the same.
Therefore, when $\lambda_{0}$ increases, all the resonator modes undergo the
super-radiant transition at the same critical point $\lambda_{0}^c$. In  Fig.
\ref{Fig:phimin}(a) and (b), the results are obtained by considering only one
mode in the calculation as shown in subsection \ref{Subsect:mean_field}. Since
the plots indicate that all resonator field modes can become macroscopically
occupied at the same time, a more rigorous treatment requires including all
resonator modes in the calculation. This is challenging numerically since the
amount of calculation required increases dramatically with the number of
resonator modes included. In Fig. \ref{Fig:phimin}(c), the results are plotted
when the first 3 resonator modes are considered simultaneously. We see that,
the average values for all resonator field modes indeed become nonezero at the
same critical point, consistent with the results in Fig. \ref{Fig:phimin}(a).

For our studies, we wish to pick a particular mode to undergo the super-radiant
phase transition. This can be accomplished by making the critical value
$\lambda_0^c$ for the chosen mode lower than that of other modes.
For this purpose, we make the Ising interaction strength $J(j)$ inhomogeneous
and position dependent. We find that, by giving $J(j)$ a spatial modulation as
simple as a rectangular wave, we can lower the critical value
$\lambda_0^c$
for one particular resonator mode below that of all others.
For example,
if the position dependence of $J(j)$ is
\begin{equation}\label{eq:J1}
J(j)=
\begin{cases}
J_{max} & \text{$j\in{[N/8,3N/8]}$ and $[5N/8,7N/8]$}\\
J_{min} & \text{all other sites},
\end{cases}
\end{equation}
the critical value $\lambda_0^c$ for the mode $l=2$ is the lowest, as shown
in Fig. \ref{Fig:phimin} (d). If instead the position dependence of $J(j)$ is
\begin{equation}\label{eq:J2}
J(j)=
\begin{cases}
J_{max} & \text{$j\in{[N/12,3N/12]}$ and $[5N/12,7N/12]$}\\
      &\text{and $[9N/12,11N/12]$}\\
J_{min} & \text{all other sites},
\end{cases}
\end{equation}
then the mode $l=3$ becomes the first one to undergo the super-radiant phase
transition when $\lambda_0$ increases, as shown in Fig. \ref{Fig:phimin}
(e). In these examples, notice that the period of $J(j)$ is the same with that
of the chosen resonator mode. Also, the gap between the critical value of
$\lambda_0^c$ for the chosen mode and others increases with the amplitude of the
Ising interaction strength modulation, $\Delta J=|J_{max}-J_{min}|$. In Fig.
\ref{Fig:phimin}(d) and Fig. \ref{Fig:phimin}(e), only one mode is included in
each calculation. To check the validity of the conclusion derived from this
simplification, we also performed the calculation by including all three modes
simultaneously and plot the results in Fig. \ref{Fig:phimin}(f). It is
seen that, the critical value of $\lambda_0$ for the first mode to undergo the
super-radiant phase transition remains approximately the same with that in Fig.
\ref{Fig:phimin}(d), and there is a clear gap in the values of $\lambda_0$ for
other modes to undergo the phase transition. Therefore, by using this technique
we can in principle single out a resonator mode to undergo the super-radiant
phase transition while all other modes remain unoccupied.

\subsection{First and second order quantum phase transition}

\begin{figure}
\includegraphics[width=1.0\linewidth]{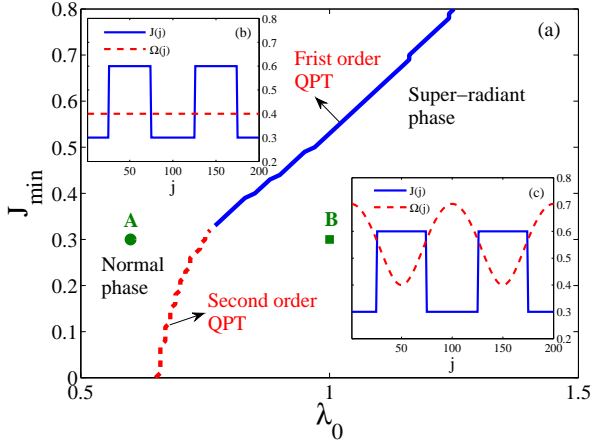}
\caption{(Color online) Phase diagram of the super-radiant quantum phase transition (QPT) in
the $J_{min}-\lambda_0$ plane.
When the phase transition occurs, the mean field value $\phi^g_2$ changes from 0
(the normal phase) to nonezero (the super-radiant phase).
When $J_{min}$ is small, the change of
$\phi^g_2$ from $0$ to nonzero is continuous. When $J_{min}$ is greater than
\textbf{$0.35$} (roughly), this change becomes discontinuous. Correspondingly,
the phase transition changes from second order to first order. In the upper
left and lower right corner, the Ising interaction strength $J(j)$ and
effective transverse field $\Omega(j)$ at point A and B in the phase diagram are
plotted. $J(j)$ is modulated as in Eq. (\ref{eq:J1}) to single out the mode
$l=2$ for the super-radiant phase transition. The effective transverse field
$\Omega(j)=\sqrt{(\frac{E_{z}}{2})^2+[2\lambda_2(j)\phi_2^g]^2}$
(see Appendix B(4)) is dependent on the order parameter $\phi_2^g$. At point A
in the normal phase, $\phi_2^g=0$ and $\Omega(j)$ is constant. At point B in the
super-radiant phase, $\phi_2^g$ is nonezero and $\Omega(j)$ is oscillatory
because of the position dependence of $\lambda_2$ as in Eq.
(\ref{eq:lambda_pos_dep}).
Parameters used in the simulation are $\frac{e^2}{2C_\Sigma}=8$, $E_{z}=0.8$,
$l=2$, $N=200$, and $\Delta J=0.3$. }

\label{Fig:phasediag}
\end{figure}

The exact nature of the quantum phase transition of the resonator field and its
relation with the Ising interaction strength $J(j)$ and transverse field $E_z$ is an interesting topic in
our problem.
To study it, we first choose a single mode to undergo the
super-radiant phase transition while all other modes remain in the unpopulated state.
Specifically, we focus on the mode $l=2$ by modulating the Ising
interaction strength as in Eq. (\ref{eq:J1}).
Assuming a transverse field $E_z=0.8$, we fix
the amplitude of the Ising-interaction modulation by setting $\Delta J=0.375 E_z=0.3$,
and use the value of $J_{min}$ as
the measure for the strength of the Ising interaction.
We then calculate the order parameter $\phi^g_2$ as a function of $J_{min}$ and the
qubit-field coupling strength $\lambda_0$. This will allow us to examine the
phase transition in great detail and determine its exact nature. The results are
plotted in Fig. \ref{Fig:phasediag}.

We find that, when the Ising interaction is weak and
the value of $J_{min}$ is small, the transition of $\phi^g_2$ from $0$ to a
nonzero value is continuous. This smooth increase in $\phi^g_2$ is the most
conspicuous signature for a second order QPT which is represented by the red
dashed line in Fig. \ref{Fig:phasediag}. On the other hand, when $J_{min}$
increases above about $0.35$, the transition to a nonzero
value for $\phi^g_2$ becomes discontinuous, indicating that the phase transition
has changed to first order. This is labeled by the blue solid line. Since the
first order phase transition grows out of a second order one, there will be a
region where the jump of $\phi^g_2$ is small \cite{C.N.,S.K.}.

\begin{figure}
\includegraphics[width=1.0\linewidth]{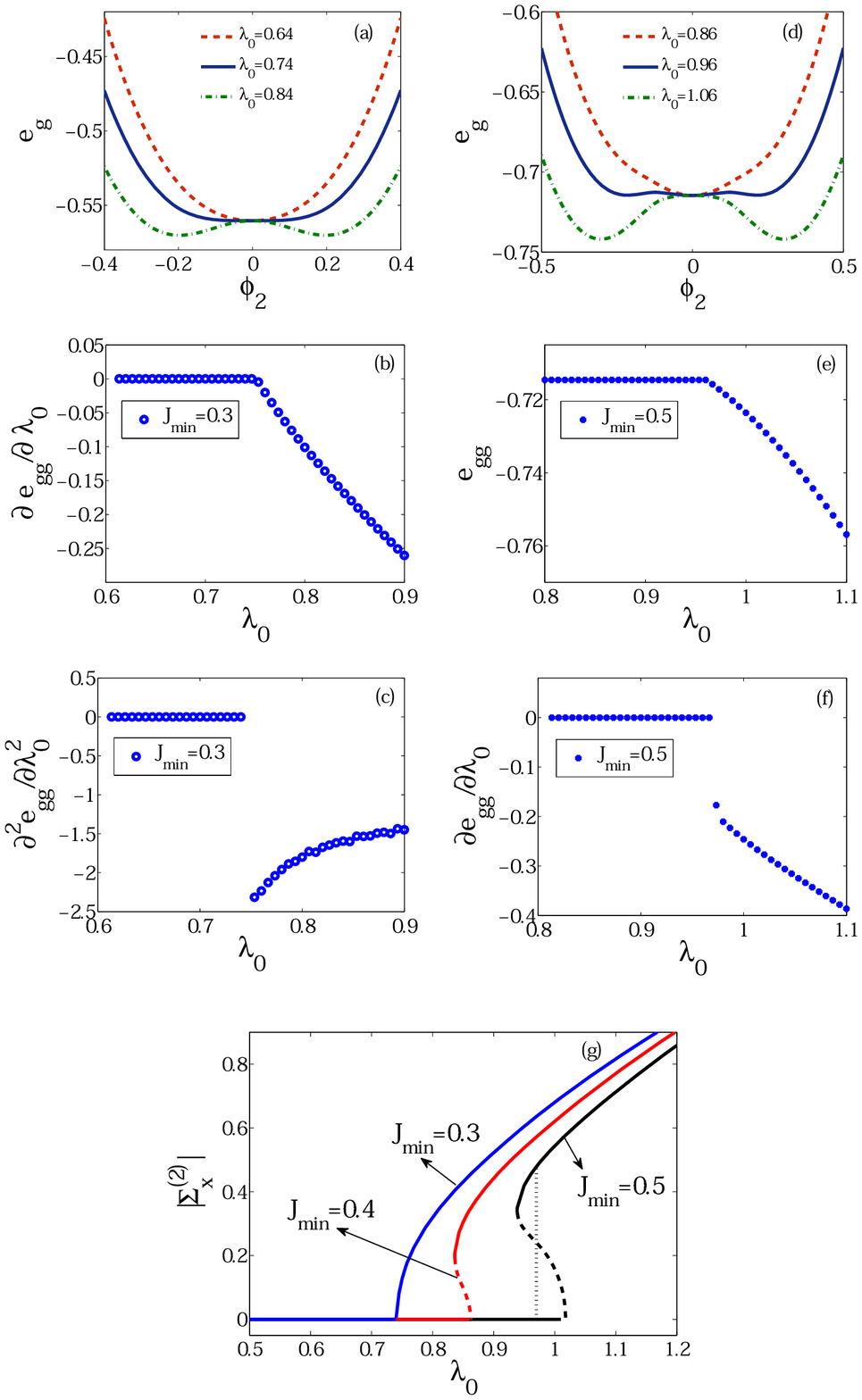}
\caption{(Color online)
(a-f) The single particle energy $e_g$, ground state energy $e_{gg}$ and the
first and second derivative of $e_{gg}$. In (a-c), $J_{min}=0.3$, and the
phase transition is second order. In (d-f), $J_{min}=0.5$, and the phase
transition is first order. Other parameters are
the same as in Fig. \ref{Fig:phasediag}. (g) $|\Sigma^x_2|$ as a function of
$\lambda_0$
for different values of $J_{min}$. The red, blue, and black solid lines refer
to the values of $|\Sigma^x_2|$ for the stable points of the minimum energy in
the
$e_g-\phi_2$ plane. The dashed lines refer to the values of $|\Sigma^x_2|$ for
the
unstable points of the maximum energy in the $e_2-\phi_2$ plane. The vertical
dotted line marks the critical point of the first order QPT for
$J_{min}=0.5$. To the left of the
dotted line, $|\Sigma^x_2|=0$ for the ground state of the system.
To the right of this line, the value of $|\Sigma^x_2|$ for the ground state
becomes nonzero.}
\label{Fig:QPT order}
\end{figure}

To demonstrate clearly the differences between the second and first order
QPT, we consider two cases where $J_{min}$ is much smaller and much greater
than $0.35$, and calculate the single particle energy $e_g$ as a function
of $\phi_2$ for different values of $\lambda_0$ near the critical point
$\lambda_0^c$. This will reveal how the strength of the Ising interaction
$J_{min}$ impacts the nature of the phase transition. The result for
$J_{min}=0.3$ is shown in Fig. \ref{Fig:QPT order}(a). In the curves for
$e_g$, we see that the single minimum at zero field continuously splits
into two symmetrically located
minima as the field-qubit coupling $\lambda_0$ is increased. This smooth
transition signals a second order QPT. To verify this, we further
calculate the first and second derivative of the ground state energy $e_{gg}$
with respect to the parameter $\lambda_0$ and plot the result in Figs.
\ref{Fig:QPT order} (b) and (c) \cite{K.B.,P.B.}. It is seen that the first
derivative is continuous, whereas the second derivative is discontinuous. We
can then conclude that the phase transition is indeed second order in this
case. In Fig. \ref{Fig:QPT order}(d), the single particle energy with
$J_{min}=0.5$ is shown for different values of $\lambda_0$. In these curves, as
$\lambda_0$ increases, the number of local minima in $e_g$ changes from one to
three and then to two. When the two minima at nonezero $\phi^g_2$ appears, the
original local minimum at 0 field does not vanish and remains the global minimum
of the system. When $\lambda_0$ increases further, the energy at the local
minima corresponding to nonzero $\phi^g_2$ abruptly become the global minimum.
In Figs. \ref{Fig:QPT order} (e) and (f), the ground state energy and
its first derivative with respect to $\lambda_0$ are plotted. Since the first
derivative is discontinuous, the phase transition in this case is first
order.

The reason for the QPT changing to first order is that, when the Ising
interaction is strong, the qubit chain is in the ferromagnetic phase before the
super-radiant phase transition occurs. Once the super-radiant phase
transition occurs, the qubit chain experiences a large effective transverse
field due to the nonzero field value $\phi^g_2$. As a consequence, the qubit
chain may abruptly switch to a paramagnetic phase, which in turn leads to a
discontinuous change in the first order derivative of $e_{gg}$.

We can further study the nature of the phase transition by investigating the
magnetic properties of the qubit chain. In Appendix B, we show that, at the
minima or maxima of $e_g(\phi_2)$, the order parameter for the qubit system is
related to that of the field according to
\begin{equation}\label{eq:phi-Sx}
\phi_2=\frac{\Sigma^x_2}{\omega_2+4D_2}.
\end{equation}
In Fig. \ref{Fig:QPT order}(c), we plot $|\Sigma^x_2|$ at the
minima or maxima of $e_g$ versus $\lambda_0$. We can see
that, when $J_{min}=0.3$, $|\Sigma^x_2|$ changes continuously with
$\lambda_0$ which indicates that the QPT is second-order in nature. When
$J_{min}=0.4$ or $0.5$, the curve for $|\Sigma^x_2|$ is hysteretic,
suggesting that a first-order QPT takes place.

\begin{figure}
\includegraphics[width=1.0\linewidth]{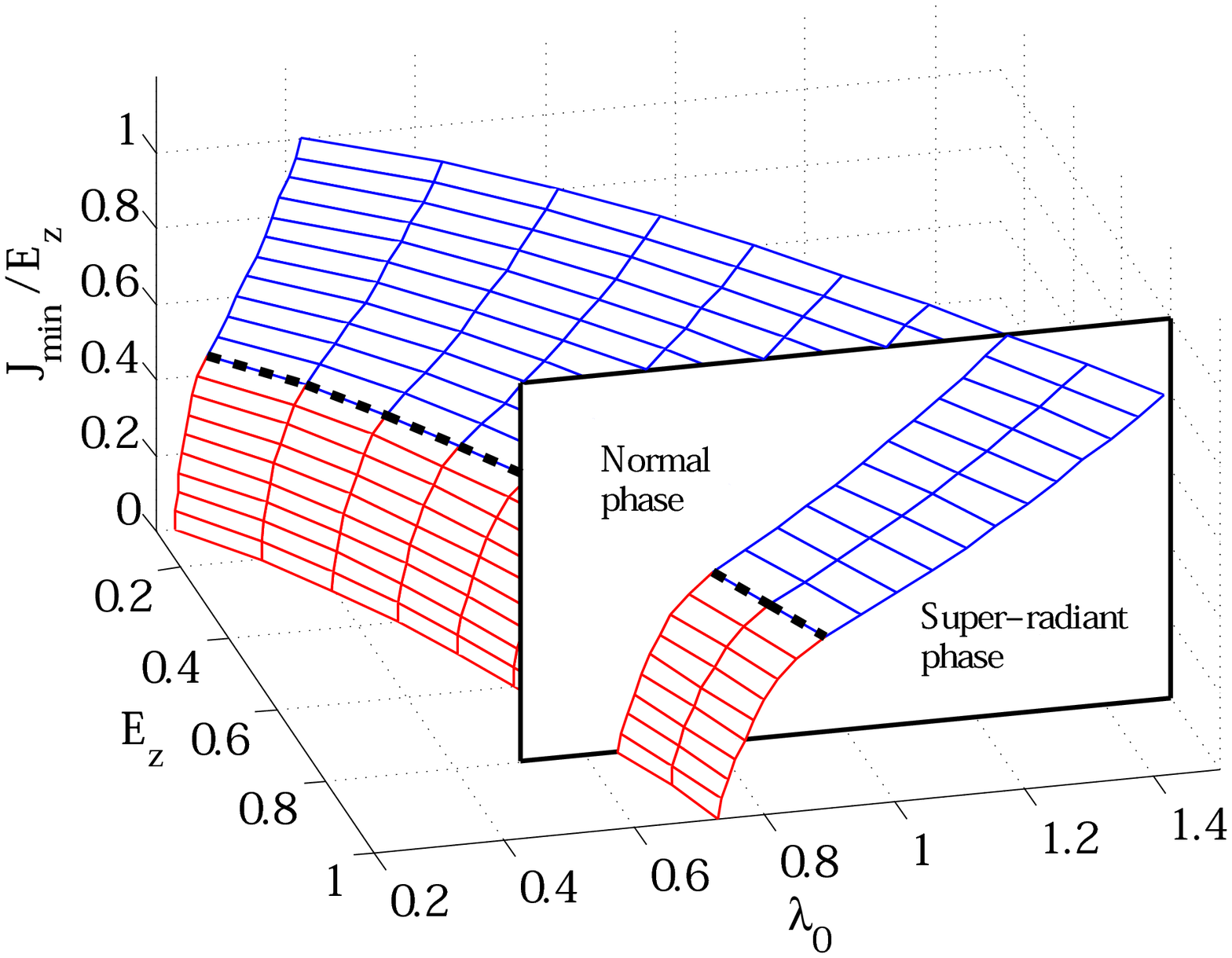}
\caption{The full phase diagram in the parameter space $(E_z, \lambda_0,
J_{min})$. The $\lambda_0-J_{min}$ plane at $E_z=0.8$ (which corresponds to Fig.
\ref{Fig:phasediag}) is shown, and the black dashed line marks the boundary
between the first and second order phase transition. The effective transverse
field is constant in the normal phase and oscillatory in the super-radiant
phase as in Fig. \ref{Fig:phasediag}. Here, we set $\Delta J=0.375E_z$.
Other parameters are the same as Fig.\ref{Fig:phasediag}.
}
\label{fig:phasediagramfull}
\end{figure}

In Fig. \ref{Fig:phasediag}, a typical value of 0.8 was used for the
transverse field $E_z$. To study the dependence of the QPT on $E_z$, we
calculate the phase diagram in the 3-dimensional parameter space $(E_z,
\lambda_0, J_{min})$ and plot the result in Fig. \ref{fig:phasediagramfull}. In
this calculation, still the $l=2$ mode is picked for the super-radiant phase
transition, and the modulating amplitude is fixed at $\Delta J=0.375E_z$. It is
seen that the basic structure of the phase diagram remains the same as in Fig.
\ref{Fig:phasediag} at different values of $E_z$, though the critical value of
$\lambda_0^c$ increases as $E_z$ grows.

\subsection{Magnetic orders in the ground states}

The qubit part of our system is essentially a nonlocal Ising chain subject to a
transverse magnetic field dependent on the state of the resonator field. In a
homogeneous Ising chain ($J$ position independent) with uniform transverse
field, the physics is dictated by the competition between the Ising interaction
and transverse field, and it is well known that the system has a critical point
when the two are equally strong. In our system, this mechanism continues to play
a major role. In addition, the state of the resonator field and its phase
transition has a nontrivial impact on the property and behavior of the qubit
chain, and we expect richer physics due to the interplay between the qubit and
resonator field.

To study the properties of the qubit chain, we focus our attention on the qubit
correlation $\langle\bar\sigma_j^y\bar\sigma^y_{j+n}\rangle$ (see Eq.
(\ref{eq:correlation}) in Appendix \ref{Appendix:correlation}), where
$\bar{\sigma}_j^y$ is the Pauli matrix of the $j$th qubit in the direction of
the Ising interaction. Since this
correlation decreases with the qubit separation $n$, we can use it to
characterize the correlation properties of the qubit
chain. For the inhomogeneous Ising chain in our problem, we define $\xi_R(j)$
the right correlation length for the $j$-th qubit if
$\langle\bar\sigma_j^y\bar\sigma^y_{j+\xi_R}\rangle=e^{-1}
\langle\bar\sigma_j^y\bar\sigma^y_{j+1}\rangle$. Likewise, we define
$\xi_L(j)$ the left correlation length for the $j$-th qubit if
$\langle\bar\sigma_j^y\bar\sigma^y_{j-\xi_L}\rangle=e^{-1}
\langle\bar\sigma_j^y\bar\sigma^y_{j-1}\rangle$.
$\xi_{RL}(j)=[\xi_R(j)+\xi_L(j)]/2$ is the average of the left and right
correlation length for the $j$-th qubit.

\begin{figure}
\includegraphics[width=1.0\linewidth]{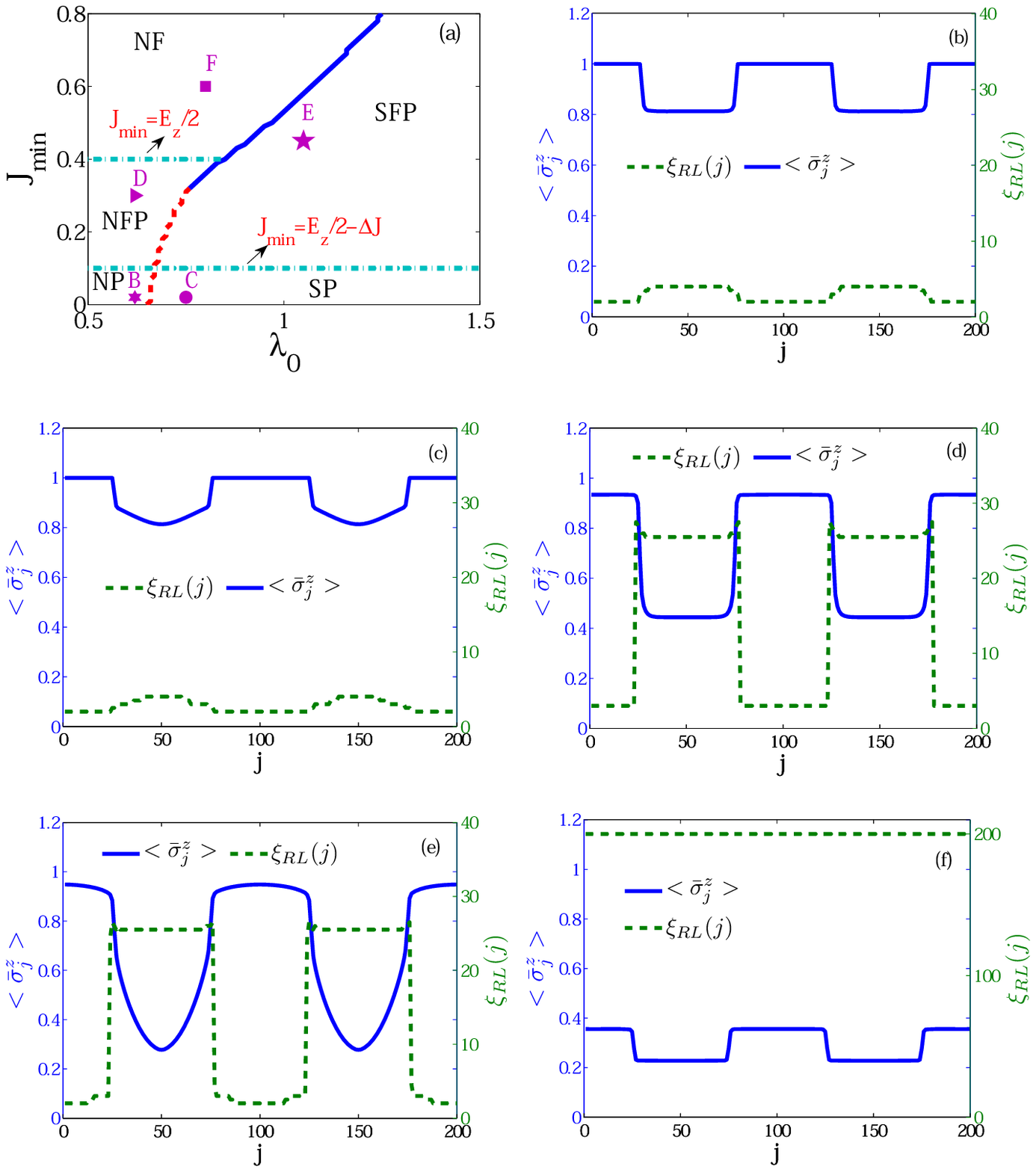}
\caption{(Color online)(a)Phase diagram of the system. All parameters are the same as in
Fig. \ref{Fig:phasediag} (a). (b-f) Values of $\langle\bar{\sigma}_j^z\rangle$
and the correlation length $\xi_{LR}(j)$ at point (B-F) in
(a).}
 \label{Fig:Isingphase}
\end{figure}

In Fig. \ref{Fig:Isingphase}, we calculate and plot the mean spin
$\bar{\sigma}_j^z$ and correlation length $\xi_{RL}(j)$ for a few representative
points in the phase space of the system, using methods developed in Appendix
\ref{Appendix:correlation}. These points are selected such that
they span both the normal and super-radiant phase of the resonator field, and
cover both the weak and strong Ising interaction regime. Information obtained
from the plots of $\bar{\sigma}_j^z$ and $\xi_{RL}(j)$ can then help us
understand the impact of the resonator field and Ising interaction on the qubit
chain.

The situation when the resonator field is in the normal phase regime with
$\phi^g_2=0$ is shown in Fig. \ref{Fig:Isingphase} (b),(d), and (f). Since the
resonator field is unpopulated, the transverse field is simply $E_z/2$, and
the state of the qubit chain is mainly determined by its competition with the
Ising interaction strength. It can be seen in Fig. \ref{Fig:Isingphase}
(b) that, when the Ising interaction is weak (roughly speaking,
$J_{min}<E_z/2-\Delta J$), the ground state of the system exhibits the
normal-paramagnetic (NP) order with a large $\langle\bar{\sigma}_j^z\rangle$ and
a small correlation length $\xi_{LR}(j)$. In contrast, when all local Ising
interaction dominates the transverse field ($J_{min}>E_z/2$), the qubit
chain is in a normal-ferromagnetic (NF) state with a small
$\langle\bar{\sigma}_j^z\rangle$
and a large correlation length $\xi_{LR}(j)$, as shown in Fig.
\ref{Fig:Isingphase} (f). In between these two cases
($J_{min}<E_z/2<J_{max}$), we have an interesting scenario where the
transverse field and Ising interaction is dominant in different segments of the
qubit chain. Consequently, both the paramagnetic and ferromagnetic orders
are present in the system. This is evidenced by the oscillating behavior of
$\langle\bar{\sigma}_j^z\rangle$ and $\xi_{LR}(j)$ along the qubit chain, as
shown in Fig. \ref{Fig:Isingphase} (d). We call it the
normal-ferromagnetic-paramagnetic (NFP) order.

When the resonator field is in the super-radiant phase regime, $\phi^g_2\neq 0$,
the effective transverse field for the $j$th qubit is position and $\phi^g_2$
dependent (see Eq. (\ref{eq:effOmeg}) in Appendix. B). If the Ising interaction
is weak,
$J_{max}<E_z/2$, the local transverse
field $\Omega(j)$ is always larger than $J(j)$ along the qubit chain. This leads
to the super-radiant-paramagnetic (SP) order with a large
$\langle\bar{\sigma}_j^z\rangle$ and a small correlation length $\xi_{LR}(j)$,
as shown in Fig. \ref{Fig:Isingphase} (c). As the strength of the Ising
interaction increases, it is possible for the transverse field to dominate
($\Omega(j)>J(j)$) in some segments and the Ising interaction to dominate
($J(j)>\Omega(j)$) in the remaining of the qubit chain. As shown in Fig.
\ref{Fig:Isingphase} (e), the ground state of the system exhibits the
super-radiant-ferromagnetic-paramagnetic (SFP) order characterized by
oscillating $\langle\bar{\sigma}_j^z\rangle$ and $\xi_{LR}(j)$. For the
parameters we calculated, there is no super-radiant-ferromagnetic
(SF) order when the strength of the Ising interaction is increased further,
and $\phi^g_2$ coexists with $\langle\bar{\sigma}_j^z\rangle$.

\section{experimental consideration}

In solving the model Hamiltonian in Eq. (\ref{eq:Ht}) and investigating
possible phase transitions in the system, we have explored a large range for
the values of relevant parameters in the model. In reality, the reachable
parameter space is limited by the currently available technology. For the charge
box - TLR system in Fig. \ref{Fig:model}, the TLR frequency and the Josephson
energy of the charge boxes are typically around a few Gigahertz. The
coupling strength between a single qubit and the TLR field can range from a few
KHZ to nearly $1$GHz \cite{H.H., W.S.}. Since the effective
coupling strength $\lambda_0$ is proportional to $\sqrt{N}$, a larger number of
charge boxes placed in the TLR will result in a stronger coupling. However,
the number of qubits in the charge box array is limited by decoherence and the
requirement for the two-state approximation to hold \cite{Du}.

\begin{figure}
\includegraphics[width=1.0\linewidth]{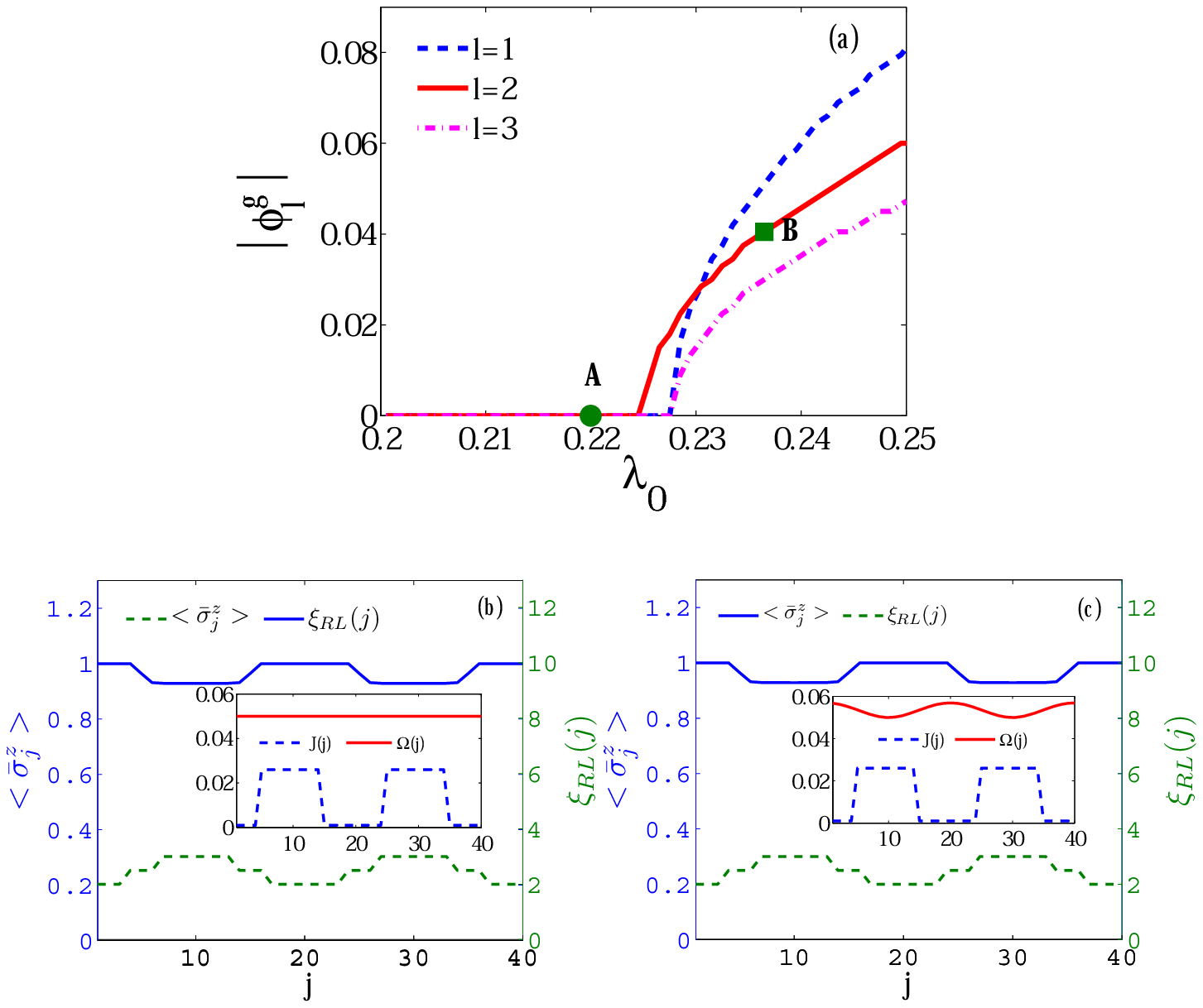}
\caption{(Color online) (a)$|\phi_l^g|$ versus $\lambda_0$ for different values of $l$ with
$E_z=0.1$ (in unit of $\omega_1$), and $J_{max}=0.26E_z$, $J_{min}=0.01E_z$. In the normal phase,
the order parameter $\phi_l^g=0$. As $\lambda_0$ crosses the critical value,
$\phi_l^g$ becomes nonzero, and the system enters super-radiant phase. (b) and
(c) are $<\bar{\sigma}^z(j)>$ and $\xi_{RL}(j)$ for the normal phase (point A)
and
super-radiant phase (point B). Both are in the paramagnetic regime.}
\label{fig:N40}
\end{figure}

Fig. \ref{fig:N40} shows our numerical results with the following parameters:
$\omega_1\simeq3\text{GHz}$, the $2nd$ mode frequency
$\omega_2=2\omega_1$, $E_z=0.1\omega_1$, $N=40$ and
$\lambda_0\in[0,0.25\omega_1]$. These parameters are accessible in present
experiments \cite{A.B., W.S., R.J.}. Considering the challenge in realizing very strong
Ising interaction \cite{C.A.}, we use experimentally accessible values
$J_{max}=0.26E_z$ and $J_{min}=0.01E_z$ \cite{C.A.,D.C.}. As
an example, we consider the case where the Ising interaction strength is
modulated according to Eq. (\ref{eq:J1}) and the second resonator mode
undergoes the super-radiant transition.
Since $J(j)<0.26E_z$ is in the weak interaction regime, the qubit chain
is restricted to the NP and SP phases. Plotted in Fig. \ref{fig:N40}(b) and
(c) are the correlation lengths and $\langle\bar{\sigma}_j^z\rangle$
at point A (in the normal phase) and B (in the super-radiant phase) in
Fig. \ref{fig:N40}(a). Because of
the limited Ising interaction strength, the main characteristics of these two
plots are similar. To achieve the NFP and SFP phases, stronger Ising
interaction strength
is needed which is still challenging experimentally.

\section{summary}

In summary, we have studied phase transitions in an Ising chain with
transverse field and coupled to a multi-mode resonator field beyond the
long-wavelength approximation. We find that the super-radiant phase transition
occurs when the coupling between the qubit and resonator
field is strong enough, and we show that we can pick a particular field mode to
undergo the super-radiant phase transition by properly modulating the Ising
interaction strength. We further studied the magnetic properties of the Ising
chain, and discovered a rich set of possible phases by calculating the qubit
correlation functions.

\section{Acknowledgement}
This work was funded by National Natural Science Foundation of
China (Grant No. 11174270), National Basic
Research Program of China 2011CB921204, 2011CBA00200, the Strategic Priority Research
Program of the Chinese Academy of Sciences (Grant No.
XDB01000000) and Research
Fund for the Doctoral Program of Higher Education of China (Grant
No. 20103402110024). Z. -W. Zhou gratefully acknowledges the
support of the K. C. Wong Education Foundation, Hong Kong.

\appendix

\section{Hamiltonian of the charge qubit chain - TLR
system}
\label{Appendix:Hamiltonian}

We derive the Hamiltonian for the circuit in Fig. \ref{Fig:model} in this
section. First, the quantized TLR modes are described by the Hamiltonian
($\hbar=1$)
\begin{equation}
H_{R}=\sum_{l}\omega_{l}b^{\dag}_lb_l,
\end{equation}
where $b_l^{(\dag)}$ is the annihilation (creation) operator for the $l$th mode,
and $\omega_l$ is its frequency. $\omega_l$ is determined by the physical
parameters of the TLR, $\omega_l=l\pi/d\sqrt{L_0C_0}$, where $d$ is the
length of the TLR, and $C_0$ and $L_0$ are the capacitance and inductance per
unit length. The voltage of the resonator associated with the $l$th mode can be
expressed as
$V_{l}(x)=\sqrt{\frac{\omega_l}{dC_0}}\cos(k_{l}x)(b_{l}+b^\dag_{l})$, where
$k_l =\omega_l\sqrt{C_0L_0}$ is its wave vector,  and $x\in [0,d]$ is the
position along the TLR\cite{A.B.}.

Now we consider a chain of $N$ equally spaced Cooper pair boxes embedded in a
TLR \cite{A.B., A.J., P.C.} as shown in Fig. \ref{Fig:model}. Because of the
capacitive coupling to the TLR, the total gate voltage for the $j$-th qubit is
the sum of a DC bias value and a quantum part due to the TLR voltage. Therefore,
the total gate voltage is $V_g(x_j)=V_g^{dc}+\hat{V}(x_j)$, with the
quantum part $\hat{V}(x_j)=\sum_l\hat{V}_l(x_j)$. The Hamiltonian for the
charge boxes\cite{P.C.} reads
\begin{equation}
\begin{split}
H_{C}=&\sum_j\{4E_c\sum_n (n-n_g^j)^2|n\rangle_j\langle n| \\
&- E_z/2\sum_n(|n+1\rangle_j\langle n|+|n\rangle_j\langle n+1|)\},
\end{split}
\label{Eq:Qubit}
\end{equation}
where $E_c=e^2/2C_\Sigma=e^2/2(C_g+C_J)$ is the charging energy ($C_\Sigma$,
$C_g$ and $C_J$ the total, gate, and Josephson junction capacitance of the
charge box), $E_z$ is the Josephson energy, and $n_g^j=C_gV_g(x_j)/2e$ is the
excess charge on the $j$-th Cooper pair box.
If we bias the charge boxes at the degeneracy point
$C_gV_g^{dc}/2e=1/2$, the charge boxes effectively function as two
level qubits with the charge states $n=0,1$. In this case, the excess charge $n_g^j=1/2+C_g\hat{V}(x_j)/2e$.
 In the subspace $\{n=0,1\}$, the qubit charging energy is
\begin{equation}
\begin{split}
&4E_c\sum_j\sum_{n=0,1}(n-n_g^j)^2|n\rangle_j\langle n|\\
&=4E_c\sum_j\sum_{n=0,1}\big[n-\frac{1}{2}-C_g\hat{V}(x_j)/2e\big]^2|n\rangle_j\langle n|\\
&=4E_c\sum_j\sum_{n=0,1}\Big[(n-\frac{1}{2})^2-2(n-\frac{1}{2})C_g\hat{V}(x_j)/2e\\
&+(C_g\hat{V}(x_j)/2e)^2\Big]|n\rangle_j\langle n|\\
&=\sum_j\Big[E_c-4E_c\frac{C_g\hat{V}(x_j)}{2e}(|1\rangle_j\langle 1|-|0\rangle_j\langle 0|)\\
&+4E_c\frac{C_g^2\hat{V}^2(x_j)}{4e^2}\Big]
\end{split}
\label{eq:charge}
\end{equation}
where we have used $|1\rangle_j\langle 1|+|0\rangle_j\langle 0|=1$. Recall that
\begin{equation}
\hat{V}(x_j)=\sum_l\sqrt{\frac{\omega_l}{dC_0}}\cos(k_{l}x_j)(b_{l}+b^\dag_{l})
\end{equation}
The second term in Eq. (\ref{eq:charge}) is the coupling between the TLR and the qubit
\begin{equation}
\begin{split}
H_{Q-R}=4E_c\frac{C_g}{2e}\sum_j\sum_l\Big[\sqrt{\frac{\omega_l}{dC_0}}\cos(k_{l}x_j)\\
(b_{l}+b^\dag_{l})(|1\rangle_j\langle 1|-|0\rangle_j\langle 0|)\Big].
\end{split}
\end{equation}
The third term  gives rise to the self energy of the TLR
\begin{equation}
H_{R-R}=4E_c\frac{C_g^2}{4e^2}\sum_j\sum_l{\frac{\omega_l}{dC_0}}\cos^2(k_{l}x_j)(b_{l}+b^\dag_{l})^2,
\end{equation}
where we have used $\sum_j\cos(k_lx_j)\cos(k_{l'}x_j)\sim0$ for $l\neq l'$ and ignored the coupling between different resonator modes.
The Hamiltonian of the qubits is given by the Josephson energy term in Eq. (\ref{Eq:Qubit})
\begin{equation}
H_Q=-\frac{E_z}{2}\sum_j(|1\rangle_j\langle 0|+|0\rangle_j\langle 1|).
\end{equation}
In the qubit eigenstates $\{(|0\rangle\pm |1\rangle)/\sqrt{2}\}$,
the total Hamiltonian of the system then reads
\begin{equation}
\begin{split}
&H=H_{R}+H_Q+H_{Q-R}+H_{R-R},\\
&H_Q=-\frac{E_z}{2}\sum_j\sigma_j^z,\\
&H_{Q-R}=-\sum_j\sum_l\frac{\lambda_l(j)}{\sqrt{N}}\sigma_j^x(b_l^\dag+b_l),\\
&H_{R-R}=\sum_lD_l(b_l+b_l^\dag)^2,
\end{split}
\label{Eq:H_no_QQ}
\end{equation}
where $\lambda_l(j)=\lambda_0\sqrt{l}\cos(l{\pi}j/N)$,
$D_l=\frac{1}{N}\sum_j\frac{C_\Sigma}{2e^2}\lambda_l^2(j)$, with
$\lambda_0=\frac{eC_g}{C_{\Sigma}}\sqrt{\frac{N\omega_1}{dC_0}}$.

Further, adjacent charge boxes can be coupled using an rf-SQUID mediated
tunable coupler as shown in Fig. (\ref{Fig:model}). The rf-SQUID acts as an
inductive transformer leading to an effective mutual inductive energy
\cite{B.B., T.A.}.
\begin{equation}
H_{Q-Q}=-\sum_jM_{eff}(j)I_jI_{j+1},
\end{equation}
where $M_{eff}(j)$ is the effective mutual inductance, and $I_j$ is the total
current through the $j$-th  junction.
For charge qubits, we have $I_j=-\frac{C_g}{2e}\ddot{\varphi}_j$ and
$\frac{C_\Sigma}{2e}\ddot{\varphi}_j+I_c\text{sin}
(\varphi_j) = 0$, where $\varphi_j$ is the junction phase,
and $I_c$ is
the critical current. Then we have
$I_j=\frac{C_g}{C_{\Sigma}}I_c\text{sin}(\varphi_j)$ and
\begin{equation}
\begin{split}
H_{Q-Q}=&-\sum_jJ(j)\text{sin}(\varphi_j)\text{sin}(\varphi_{j+1})\\
=&-\sum_jJ(j)\sigma_j^y\sigma_{j+1}^y
\end{split}
\label{Eq:H_QQ}
\end{equation}
where $J(j)=\frac{M_{eff}(j)C_g^2I_c^2}{C_{\Sigma}^2}$, and we have used
$\text{sin}(\varphi)=\sigma^y$. It is assumed that the coupler is placed far
away
from the TLR and the coupling to TLR can be ignored.

In summary, the total Hamiltonian of the system is
\begin{equation}
 H=H_{R}+H_Q+H_{Q-R}+H_{R-R}+H_{Q-Q},
 \label{Eq:H_tot}
\end{equation}
where each component of the Hamiltonian is given in Eq. (\ref{Eq:H_no_QQ}) and
(\ref{Eq:H_QQ}).

\section{Mean field solution}
\label{Appendix:JW}
The Hamiltonian in Eq. (\ref{Eq:H_tot}) can be solved in the mean field
approximation for the resonator field. For simplicity of presentation, we
consider only one single TLR mode $l$ first. Under the mean field
approximation for $H_{Q-R}$ as in Eq.
(\ref{Eq:mean_field_coupling}), the total Hamiltonian reads
\begin{equation}
\begin{split}
H=&\omega_l b_l^{\dag}b_l+D_l(b^{\dag}_l+b_l)^2-\sqrt{N}\Sigma^x_l(b_l^{\dag}+b_l)\\
&-\sum_j\frac{E_z}{2}
\sigma_j^z-\sum_j2\lambda_l(j)\phi_l\sigma_j^x-\sum_jJ(j)\sigma_j^y\sigma_{j+1}
^y \\
&+2N\phi_l \Sigma_l^x,
\end{split}
\label{Eq:H_mean_field}
\end{equation}
where $\phi_l$ and $\Sigma^x_l$ are the mean values of the resonator field
and qubit chain. The first line of Eq. (\ref{Eq:H_mean_field}) can be
diagonalized, and the Hamiltonian then becomes
\begin{equation}
\begin{split}
H=&\bar{\omega_l}\bar{b}_l^{\dag}\bar{b}_l-\frac{N(\Sigma^x_l)^2}{\bar{\omega}_l
(\alpha+\beta)^2}\\
&-\sum_j\frac{E_z}{2}
\sigma_j^z-\sum_j2\lambda_l(j)\phi_l\sigma_j^x-\sum_jJ(j)\sigma_j^y\sigma_{j+1}
^y\\
&+2N\phi_l \Sigma_l^x,
\end{split}
\end{equation}
where $\bar{b}_l=\alpha b_l+\beta
b^{\dag}_l+\frac{\sqrt{N}\Sigma^x_l}{\bar{\omega}_l(\alpha+\beta)}$ with
$\bar{\omega}_l=\sqrt{\omega_l^2+4D_l\omega_l}$,
$\alpha=\sqrt{\frac{\omega_l+2D_l+\bar{
\omega}_l}{2\omega_l}}$,
and $\beta=\sqrt{\frac{\omega_l+2D_l-\bar{\omega}_l}{2\omega_l}}$.

The second line of Eq. (\ref{Eq:H_mean_field}) describes a nonlocal Ising chain
with nonuniform transverse field. To find its spectrum, we first make a local
rotation along the $y$ axis to introduce the Pauli matrices
$\bar{\sigma}^z_j=\text{cos}(\theta_j)\sigma_j^z+\text{sin}(\theta_j)
\sigma_j^x$,
$\bar{\sigma}^x_j=\text{cos}(\theta_j)\sigma_j^x-\text{sin}(\theta_j)
\sigma_j^z$, and $\bar{\sigma}_j^y=\sigma^y_j$, with
$\theta_j=\text{arctan}\bigg
[\frac{4\lambda_l(j)\phi_l}{E_z}\bigg]$. Then the second line of Eq.
(\ref{Eq:H_mean_field}) takes the form
\begin{equation}
\begin{split}
H_{\text{Ising}}=-\sum_j\Omega(j)\bar{\sigma}
_j^z-\sum_jJ(j)\bar\sigma_j^y\bar\sigma_{j+1}^y,\\
\end{split}
\label{eq:Ising}
\end{equation}
where the effective transverse magnetic field
\begin{equation}\label{eq:effOmeg}
\Omega(j)=\sqrt{(\frac{E_{z}}{2})^2+[2\lambda_l(j)\phi_l]^2}.
\end{equation}
We assume the periodic boundary condition for the qubit chain,
$\bar{\sigma}_{N+1}=\bar{\sigma}_{1}$. Following the method given in \cite{A.H.,
O.D., O.J., S.S., J.D., E.T.}, we express the Pauli matrices using the creation
and annihilation operators in the spinor space
\begin{equation}
\begin{split}
\bar\sigma_{j}^z=&1-2a^\dag_{j}a_{j},\\
\bar\sigma_{j}^y=&a^\dag_{j}+a_{j},
\end{split}
\label{eq:JW}
\end{equation}
and apply the Jordan-Wigner transformation
\begin{equation}
\begin{split}
a^\dag_{j}=c^\dag_{j}e^{-i\pi\sum_{i=1}^{j-1}c^\dag_{i}c_i},\\
a_{j}=e^{-i\pi\sum_{i=1}^{j-1}c^\dag_{i}c_i}c_{j},
\end{split}
\end{equation}
to map the qubit chain to a collection of fermions described by the creation and
annihilation operators $c^\dag$ and $c$ which satisfy $\{c_{i},
c_{j}^\dag\}=\delta_{ij}$ and
$\{c_{i}, c_{j}\}=\{c_{i}^{\dag}, c_{j}^\dag\}=0$. After this transformation,
we obtain a quadratic Hamiltonian in fermion operators
\begin{equation}\label{eq:fermi}
\begin{split}
H_{\text{Ising}}=&-\sum_{j=1}^N\Omega(j)(1-2c^\dag_{j}c_j)\\
&-\sum_{j=1}^{N-1}J(j)(c_{j}^\dag-c_{j})(c_{j+1}^\dag+c_{j+1})\\
&+J_N(c_{N}^\dag-c_{N})(c_{1}^\dag+c_{1})e^{i\pi\mathcal{N}}\\
=&\sum_{i,j}c_{i}^\dag{A_{ij}}c_{j}+\sum_{i,j}\big(c_{i}^\dag{B_{ij}}c_{j}
^\dag+c_{i}B_{ij}c_{j}\big)
\end{split}
\end{equation}
where
\begin{equation}
\begin{split}
&A_{j,j}=\Omega_j, A_{j,j+1}=-\frac{J(j)}{2}, A_{j+1,j}=-\frac{J(j)}{2}, \\
&B_{j,j+1}=\frac{J(j)}{2}, B_{j+1,j}=-\frac{J(j)}{2}, \\
&A_{N,1}=-\frac{J(N)}{2}\cdot e^{i\pi\mathcal{N}}, A_{1,N}=-\frac{J(N)}{2}\cdot
e^{i\pi\mathcal{N}},\\
&B_{N,1}=\frac{J(N)}{2} \cdot e^{i\pi\mathcal{N}}, B_{1,N}=-\frac{J(N)}{2}\cdot
e^{i\pi\mathcal{N}},
\end{split}
\label{Eq:fermion}
\end{equation}
and $\mathcal{N}=\sum_{j=1}^{N}c^\dag_{j}c_j$ is the number of fermions. Though
the spin problem has a periodic boundary condition, the transformed fermion
problem could have a periodic or antiperiodic boundary condition, depending on
the fermion number $\mathcal{N}$. Specifically, the fermion problem has an
antiperiodic boundary condition if there is an even number of fermions, and
periodic boundary condition if there is an odd number of fermions. The ground
state is in the sector with antiperiodic boundary condition \cite{A.H.}.

The bilinear Hamiltonian in Eq. (\ref{Eq:fermion}) can be diagonalized exactly.
To do so, we perform the linear canonical transformation 
\begin{equation}
\eta_{k}=\sum_{j}\big(g_{kj}c_{j}+h_{kj}c_{j}^\dag\big)
\label{eq:lct1}
\end{equation}
\begin{equation}
\eta_{k}^\dag=\sum_{j}\big(g_{kj}c_{j}^\dag+h_{kj}c_{j}\big)
\label{eq:lct2}
\end{equation}
where $\eta_k$'s are a new set of fermionic quasiparticle operators, $\{\eta_k,
\eta_{k'}^\dag\}=\delta_{kk'},    \{\eta_k, \eta_{k'}\}=\{\eta_{k}^\dag,
\eta_{k'}^\dag\}=0$, and the
coefficients are chosen to be real. In order to diagonalize the Hamiltonian and
express it in the form
\begin{equation}
\begin{split}
{H}_{\text{Ising}}=\sum_{k=1}^{N}\Lambda_{k}\big(\eta_{k}^\dag\eta_{k}-\frac{1}{
2}\big),\\
\end{split}
\end{equation}
the coefficients $g_{ki}$ and $h_{ki}$ must satisfy \cite{E.T.}
\begin{align}
\Lambda_{k}^2\Phi_{k,j}=\sum_i\Phi_{k,i}(A-B)(A+B)_{i,j},\\
\Lambda_{k}^2\Psi_{k,j}=\sum_i\Psi_{k,i}(A+B)(A-B)_{i,j},
\end{align}
where $\Phi_{k,j}$ and $\Psi_{k,j}$ are linear combinations of $g_{ki}$,
$h_{ki}$,
\begin{align}
\Phi_{kj}=g_{kj}+h_{kj},\\
\Psi_{kj}=g_{kj}-h_{kj}.
\end{align}
By solving these equations, we can obtain the quasiparticle spectrum
$\Lambda_k$ and the coefficients $g_{ki}$ and $h_{ki}$.

The total Hamiltonian then reads
\begin{equation}
\begin{split}
H=&\bar{\omega_l}\bar{b}^{\dag}_l\bar{b}_l-\frac{N(\Sigma^x_l)^2}{\bar{\omega}_l
(\alpha+\beta)^2}\\
&+\sum_{k=1}^{N}\Lambda_{k}\big(\eta_{k}^\dag\eta_{k}-\frac{1}{2}\big)\\
&+2N\phi_l \Sigma^x_l.
\end{split}
\end{equation}
The ground state $|G\rangle$ must satisfy
\begin{equation}
\begin{cases}
\bar{b}_l|G\rangle&=0,  \\
\eta_k|G\rangle&=0   (\forall k),
\end{cases}
\end{equation}
and the ground state energy is
\begin{equation}
E_g=-\frac{N(\Sigma^x_l)^2}{\bar{\omega}_l(\alpha+\beta)^2}-\sum_{k=1}^{N}\frac{
1 } { 2 }
\Lambda_{k}+2N\phi_l \Sigma^x_l.
\end{equation}
From $\bar{b}_l|G\rangle=0$, we get
\begin{equation}\label{eq:phi-S}
\phi_l=\langle{G}|\frac{(b^{\dag}_l+b_l)}{2\sqrt{N}}|G\rangle=\frac{\Sigma^x_l}{
\omega_l+4D_l},
\end{equation}
and therefore
\begin{equation}
E_g=N(\omega_l+4D_l)\phi_l^2-\sum_{k=1}^{N}\frac{1}{2}\Lambda_{k}.
\end{equation}
Notice that $\Omega(j)=\sqrt{(\frac{E_{z}}{2})^2+4\lambda_l^2(j)\phi_l^2}$, and
$\Lambda_k$ is also a function of $\phi_l$.
The value of $\phi_l$ is determined by minimizing $E_g(\phi_l)$. If
$E_g(\phi_l^g)$
is the minimum, we have $\frac{\partial E_g(\phi_l)}{\partial
\phi_l}|_{\phi_l^g}=0$.
Notice
\begin{equation}
\begin{split}
&-\sum_{k=1}^{N}\frac{1}{2}\Lambda_{k}\\
&=\langle
G|-\sum_j\frac{E_z}{2}
\sigma_j^z-\sum_j2\lambda_l(j)\phi_l\sigma_j^x-\sum_jJ(j)\sigma_j^y\sigma_{j+1}
^y|G\rangle,
\end{split}
\end{equation}
and
\begin{equation}
\begin{split}
&\frac{\partial}{\partial\phi_l}\langle G|\sum_j\frac{E_z}{2}\sigma_j^z
-\sum_j2\lambda_l(j)\phi_l\sigma_j^x-\sum_jJ(j)\sigma_j^y\sigma_{j+1}
^y|G\rangle\\
&=-\langle G|\sum_j2\lambda_l(j)\sigma_j^x|G\rangle.
\end{split}
\end{equation}
Therefore, the condition $\frac{\partial E_g(\phi_l)}{\partial
\phi_l}|_{\phi_l^g}=0$
leads to
\[\Sigma^x_l=\frac{\langle
G|\sum_j\lambda_l(j)\sigma_j^x|G\rangle}{N}=\phi_l(\omega_l+4D_l),\] which is
consistent with the result in Eq. (\ref{eq:phi-S}).

We can similarly calculate the ground state energy for multiple field modes in
the thermodynamic limit. The total Hamiltonian is
\begin{equation}
\begin{split}
H=&\sum_l\big[\omega_l
b_l^{\dag}b_l+D_{l}(b_l^{\dag}+b_l)^2-\sqrt{N}\Sigma^x_l(b_l^{\dag}+b_l)\big]\\
&-\sum_j\frac{E_z}{2}\sigma_j^z-\sum_{j,l}
2\lambda_l(j)\phi_l\sigma_j^x-\sum_jJ(j)\sigma_j^y\sigma_{j+1}^y \\
&+\sum_l \big(2N\phi_l \Sigma^x_l\big),
\end{split}
\label{Eq:H_mean_field Multi}
\end{equation}
and we have $\Sigma^x_l=\phi^g_l(\omega_l+4D_l)$.
The ground state energy now reads
\begin{equation}
E_g=\sum_l[N(\omega_l+4D_l)\phi_l^2]-\sum_{k=1}^{N}\frac{1}{2}\Lambda_{k}.
\end{equation}
and $\Lambda_{k}$ is the quasiparticle spectrum of Ising chain Eq.
(\ref{eq:Ising})
with effective transverse magnetic field
\begin{equation}
\Omega(j)=\sqrt{(\frac{E_{z}}{2})^2+\big[2\sum_l\lambda_l(j)\phi_l\big]^2}.
\end{equation}
The order parameter $\phi_l^g$ is determined by minimizing $E_g(\phi_1,
\phi_2 \cdots)$.

\section{Qubit correlation function}
\label{Appendix:correlation}
Now we show how to calculate the correlation functions.
The correlation of the qubit chain at ground state can be calculated using the
Fermionic operators,
\begin{equation}\label{eq:correlation}
\begin{split}
 \rho_{j,j+n}&=\langle\bar\sigma_j^y\bar\sigma_{j+n}^y\rangle\\
             &=\Big\langle{(c_j^\dag+c_j)\big[\prod_{i=j}^{j+n-1}(c_i^\dag+c_i)(c_i^\dag-c_i)
 \big](c_{j+n}^\dag+c_{j+n})}\Big\rangle.
 \end{split}
\end{equation}
If we define
\begin{equation}
\begin{split}
C_j=c_j^{\dag}+c_j,\\
D_j=c_j^{\dag}-c_j,
\end{split}
\end{equation}
then
\begin{equation}
\rho_{j,j+n}=\langle{D_jC_{j+1}D_{j+1}\cdots{C_{j+n-1}}D_{j+n-1}C_{i+n}}\rangle.
\end{equation}
This expectation value can be evaluated by Wick's theorem \cite{S.S.,
M.D.} which relates it to a sum over products of expectation values of pairs
of operators. By making use of the inverse
transformation
\begin{equation}
\begin{split}
C_j=&\sum_k\Phi_{kj}(\eta_k^\dag+\eta_k),\\
D_j=&\sum_k\Psi_{kj}(\eta_k^\dag-\eta_k),
\end{split}
\end{equation}
and
$\eta_k|G\rangle=0$, the expectation value of any such pair is easily calculated:
\begin{equation}
\begin{split}
\langle{C_iC_j}\rangle=&\sum_{k,k'}\Phi_{ki}\Phi_{k'j}\langle{(\eta_k^\dag-\eta_k)(\eta_{k'}^\dag+\eta_{k'})}\rangle
       =\delta_{i,j},\\
\langle{D_iD_j}\rangle=&\sum_{k,k'}\Psi_{ki}\Psi_{k'j}\langle{(\eta_k^\dag-\eta_k)(\eta_k^\dag+\eta_{k'})}\rangle
       =-\delta_{i,j},\\
\langle{D_iC_j}\rangle=&\sum_{k,k'}\Psi_{ki}\Phi_{k'j}\langle{(\eta_k^\dag-\eta_k)(\eta_{k'}^\dag+\eta_{k'})}\rangle
       =-(\Psi^\top\Phi)_{ij}.
\end{split}
\end{equation}
Defining
\begin{equation}
G_{i,j}=\langle{D_iC_j}\rangle,
\end{equation}
and collecting the terms in the Wick expansion, we find
\begin{gather}
\rho_{j,j+n}=
\begin{vmatrix}
G_{j,j+1} & G_{j,j+2} & \cdots & G_{j,j+n}\\
G_{j+1,j+1} & G_{j+1,j+2} & \cdots &\vdots \\
\vdots & \vdots & \ddots& \vdots\\
G_{j+n-1,j+1} & G_{j+n-1,j+2} & \cdots & G_{j+n-1,j+n}
\end{vmatrix}.
\end{gather}
$\langle\bar\sigma_j^z\rangle$ and $\langle\bar\sigma_j^x\rangle$ can also be calculated in the same way:
\begin{equation}
\begin{split}
\langle\bar\sigma_j^z\rangle=&(\Psi^\top\Phi)_{j,j}\\
\langle\bar\sigma_j^x\rangle=&0.
\end{split}
\end{equation}
We can rotate $\bar{\bm{\sigma}}$ back to $\bm{\sigma}$,
\begin{equation}
\begin{split}
&\langle\sigma_j^y\sigma_{j+n}^y\rangle=\langle\bar\sigma_j^y\bar\sigma_{j+n}
^y\rangle,\\
&\langle\sigma_j^z\rangle=\text{cos}(\theta_j)\langle\bar\sigma_j^z\rangle,\\
&\langle\sigma_j^x\rangle=\text{sin}(\theta_j)\langle\bar\sigma_j^z\rangle.
\end{split}
\end{equation}

\end{document}